\documentclass{article}

\usepackage{arxiv}

\usepackage[utf8]{inputenc} 
\usepackage[T1]{fontenc}    

\usepackage{url} 
\usepackage{color,soul} 

\usepackage{booktabs}       
\usepackage{amsfonts}       
\usepackage{nicefrac}       
\usepackage{microtype}      
\usepackage{graphicx}

\usepackage{hyperref}
\usepackage[numbers,sort&compress]{natbib}

\usepackage{authblk}

\title{Enhancing the Extraction of Interpretable Information for Ischemic Stroke Imaging from Deep Neural Networks}

\author[1,2]{Erico Tjoa}
\author[2]{Guo Heng}
\author[1]{Lu Yuhao}
\author[1]{Cuntai Guan}

\affil[1]{School of Computer Science and Engineering, Nanyang Technological University, Singapore}
\affil[2]{Alibaba Group Holding Ltd, HealthTech Division, Hangzhou, China}

\begin{document}
\maketitle
\begin{abstract}
\par We implement a visual interpretability method Layer-wise Relevance Propagation (LRP) on top of 3D U-Net trained to perform lesion segmentation on the small dataset of multi-modal images provided by ISLES 2017 competition. We demonstrate that LRP modifications could provide more sensible visual explanations to an otherwise highly noise-skewed saliency map. We also link amplitude of modified signals to useful information content. High amplitude localized signals appear to constitute the noise that undermines the interpretability capacity of LRP. Furthermore, mathematical framework for possible analysis of function approximation is developed by analogy.
\end{abstract}

\keywords{Artificial Intelligence \and Neural Network \and Medical Imaging \and Interpretability \and Explainable AI}

\section{Introduction and Related Works}
\label{section:intro}

Deep learning (DL) artificial intelligence (AI) has demonstrated remarkable capabilities for image classification tasks in its early development. The success quickly spread to other fields and has found practical uses in the medical sector as well. For example, U-Net and its variants \cite{DBLP:journals/corr/RonnebergerFB15,DBLP:journals/corr/CicekALBR16,DBLP:journals/corr/abs-1804-03999} are used for medical image segmentation. Unfortunately, despite its great successes, DL still suffers from several short-comings, which we would like to address in this paper. Firstly, its performance is not flawless. The consequence could be immeasurable; for example, applied to autonomous vehicles, slight mishap might lead to fatality. In other cases, such as medical segmentation using multi-modal MRI scans, the best performance has yet to achieve a standard high enough to warrant confidence in the use of DL in real-life diagnosis/prognosis pipeline (see grand-challenge.org; an example of the challenge is ISLES 2017).

\par Secondly, it is well known that neural networks (NN) with DL architecture (i.e. multiple hidden layers) are treated as black-boxes. Research works focusing on explaining the inner working of NN algorithms and other machine learning methods have recently gained traction \cite{10.1145/2939672.2939778,conf_icml_KimWGCWVS18,DBLP:journals/corr/ZeilerF13}; also see review paper 
\cite{DBLP:journals/corr/abs-1907-07374}. Some interpretability methods involve the use of heat-maps delineating components of the inputs that contribute significantly to the predicted output, such as Class Activation Maps (CAM) \cite{CAM7780688} and Layer-wise Relevance Propagation (LRP) \cite{Lapuschkin2019,Samek2016InterpretingTP}; again, refer to \cite{DBLP:journals/corr/abs-1907-07374} for many of their variants. These methods are often referred to as saliency methods.

\par Many different methods have also been proposed to shed lights into the internal mechanism of DL algorithms. \cite{visualization_techreport,DBLP:journals/corr/NguyenYC16,DBLP:journals/corr/YosinskiCNFL15} use activation maximizations, showing images that are produced from mathematical optimization of neuron(s) activations. Yet others are sensitivity methods, involving the extent of changes, often gradient changes, and their effects on the output \cite{10.1145/2939672.2939778, 10.5555/1756006.1859912, DBLP:journals/corr/SelvarajuDVCPB16}. Clustering of data points \cite{DBLP:journals/corr/NguyenYC16,karpathytSNECNNN,carter2019activation} visualized under lower-dimensional spaces have also been included as part of interpretability studies. Unfortunately, the efforts have yet to achieve a ground-breaking success. Before the working mechanism of an AI can be explained, or before the AI can provide robust explanation for its own output decision, we cannot expect clinical adoption or other practical deployment of AI for fear of misdiagnosis and potentially dangerous consequences.

\par The increasing need for interpretability has become clearer nowadays. While artificial intelligence appears to be able make accurate predictions, \cite{Lapuschkin2019} shows that, using LRP, AI could produce the right answer for completely wrong reasons. \cite{ghorbani2017interpretation} shows that an imperceptible manipulation of input can completely degrade the prediction of a NN. Manipulation can be arbitrary, and \cite{Dombrowski2019ExplanationsCB} shows how easily it could be performed. Interpretability methods have to grow in capability and ease of use as fast as the vulnerabilities of AI are discovered. Otherwise, the future of AI will remain unclear. More efforts could be put into realizing regulatory frameworks such as \cite{GovernanceModelReddy} proposed to address this issue.

\par In this paper, we present our understanding of how a NN might evolve during its training phase based on empirical observation.
\begin{enumerate}
\itemsep0em 
\item NN training adjusts weights by distributing weight*input across the network where input here refers to input to an NN layer and * stands for any NN layer feed-forward operation.
\item Conversely, poorly trained network has biased weights that skew signal distributions. We can interpret it as the consequence of false local minima.
\item Low magnitude signals propagated in a NN stores significant amount of information. The magnitude is low relative to local high intensity spikes that, we believe, accumulate errors.
\end{enumerate}

\par This paper shows that the observations at least apply to LRP method which zeroes negative weights. Observations 1 and 2 thus imply the possibilities that poorly trained negative weights have been incorrectly used to offset excessive noise resulting in apparently correct prediction.

\par The problem is, poorly trained NN sometimes might still produce prediction. As a simple illustration, binary classification $f(x)\ge 0$ predicts the patient suffers from stroke and $f(x)<0$ otherwise, where x is a multi-modal brain scan. However, suppose a perfectly trained network $f$ exists, and $f(x)\approx 1$ for all brain scans of patients suffering from stroke. A poorly trained neural network $f_1 (x)$ might be still predict correctly with $f_1 (x)=10$ , and this might give rise to error to interpretability algorithm later on.

\par
\begin{figure}[ht]
\vskip 0.2in
\begin{center}
\centerline{\includegraphics[width=0.7\columnwidth, trim = {0 0cm 0 0cm}]{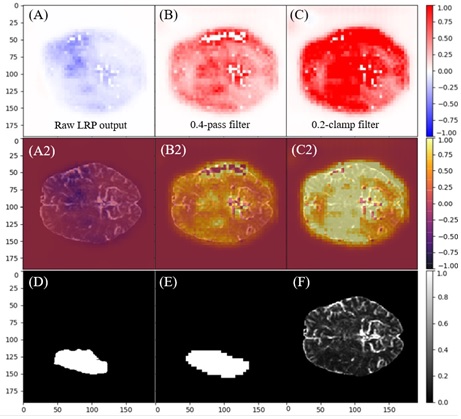}}
\caption{\textbf{Improving interpretability through more visualization of more relevant information.} LRP output for 3D U-Net A1 (see annex table 1) on patient case number 4, showing one slice of rCBF modality. Dice score is $0.769$. (A) Raw LRP output showing a large proportion of the region in the brain contributing negatively to the lesion segmentation. Suspected high amplitude noise masks away useful information. See later section. (B) LRP with $0.4$-pass filter showing the region in the brain positively contributing to the lesion segmentation. (C) LRP with $0.2$-clamp filter. (A2, B2, C2) The same LRP outputs from A, B, C superimposed on a slice in the rCBF modality of the image. (D) Lesion segmentation ground-truth. (E) Lesion segmentation provided by 3D U-Net A1. (F) A corresponding slice from rCBF modality. The signals are normalized to [-1.0,1.0] by division with the maximum absolute amplitude of each channel $max|R_c|$. The values of $max|R_c|$ for A, B and C are respectively $1241.89$, $1843.85$ and $652.63$.}
\label{fig:fig_meim3_main}
\end{center}
\vskip -0.2in
\end{figure}

 The paper is arranged as the following: (1) we demonstrate the use of LRP on 3D U-Net for ISLES 2017 challenge. (2) We show how sub-optimal LRP output can be modified to provide sensible visual explanation. By using the filters within LRP layers, we extract a more “interpretable” information. The extracted heatmap shows that LRP is using the region in the image that contains brain slice to deduce the position of the lesion. This is shown in figure \ref{fig:fig_meim3_main}(A, A2). Compare them with figure \ref{fig:fig_meim3_main}(B, B2, C, C2); also compare them with the output of LRP shown in the main LRP website \cite{tutorial:LRP}. (3) Distributions of normalized LRP output signals serve as the ground for a possible \textit{filter calculus} introduced in a later section. The technique can potentially be developed for function approximation besides finding a sweet-spot of interpretable information. Finally, (4) open source code is provided as the scaffolding for further modification \footnote{\url{https://github.com/etjoa003/medical_imaging/}. See folder isles2017.}. Note that the prediction output will not be optimal as our objective is to uncover the inner mechanism which in turn might serve as a more solid basis for network modification for both performance and interpretability improvements. 

\section{Methods and Dataset}
\par ISLES 2017 dataset is used. A small dataset of 43 patients are used as training dataset, each of which is a multi-modal image. Only 6 modalities ADC, MTT, rCBF, rCBV, Tmax and TTP are used, although clinical variables and 4D PWI are also available. The 6 modalities are prepared as 6-channel input data. Using Pytorch tensor shape notation, an example of input data has the shape $(C,D,H,W)=(6,19,192,192)$, although the sizes vary.

\par 3D U-Net is implemented with only slight modifications from the original, as shown in annex figure 1. Basic LRP implementation can be found in \cite{tutorial:LRP}. We summarize LRP here using fully-connected NN as illustration: given a prediction $y=f_{NN}(x)$ in the vector form $y=R^{(L)}$ or in component form $y_i=R_i^{(L)}$. Then,
\begin{equation}
R^{(n)}_i=\Sigma_j \frac{a_{i}^{(n)} w_{ij}^+}{\Sigma_k a_k^{(n)} w_{kj}^+} R_j^{(n+1)}
\label{eqn:lrpformula}
\end{equation}
where $w_{ij}^+$ is the weight $w_{ij}$ if it is non-negative (otherwise it is set to zero). LRP output is given by $R^{(0)}$. $L$ is the index of the last layer and layer $0$ is the input layer. For more details, also refer to \cite{SamXAI19}.

\begin{figure}[ht]
\vskip 0.2in
\begin{center}
\centerline{\includegraphics[width=0.7\columnwidth, trim = {0 0.5cm 0 0.5cm}]{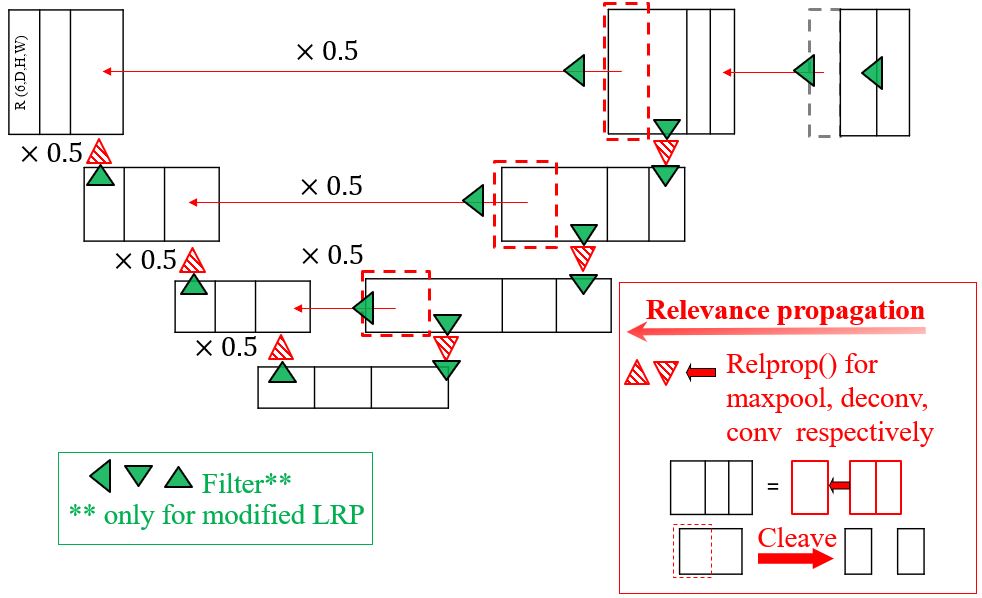}}
\caption{\textbf{LRP schematic with filters applied on 3D U-Net.} Underlying 3D U-Net architecture is shown in black, with pooling and up-convolution not shown. See annex figure 1 instead for the basic 3D U-Net architecture. Red symbols are related to LRP operations. Green symbols are filters applied for interpretability improvement.}
\label{fig:meim_unetlrp}
\end{center}
\vskip -0.2in
\end{figure}

\par LRP on 3D U-Net is implemented as shown in figure \ref{fig:meim_unetlrp}. Filters are shown in green, thus standard LRP are the same diagram without all the green symbols and with red symbols' directions reversed. The reverse of concatenation process in LRP is not defined before. In this paper, we simply cleave the feature map into two parts whose shapes are consistent with the shapes of concatenated pieces that have been forward propagated. Relevance propagation algorithm used is shown in annex algorithm 1 for applicable layers (not applicable to batch-normalization and activation layers). 

\par
\begin{figure}[ht]
\vskip 0.2in
\begin{center}
\centerline{\includegraphics[width=\columnwidth, trim = {0 0.5cm 0 0.5cm}]{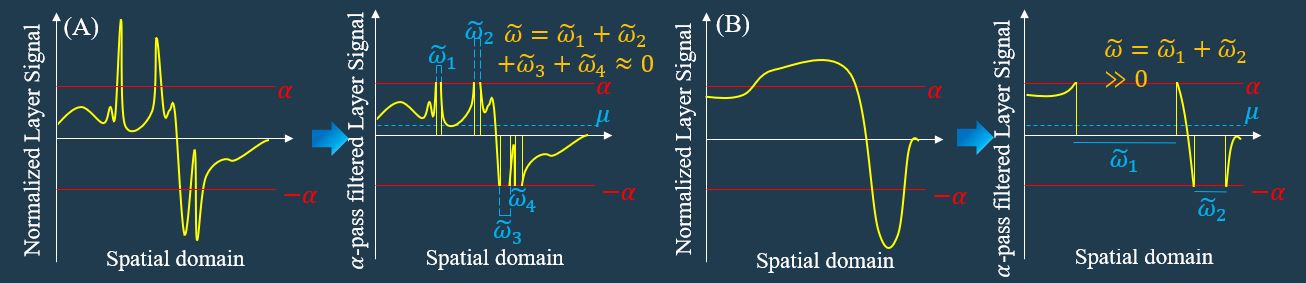}}
\caption{\textbf{$\alpha$-pass filter.} If given $\alpha$, the filter is naturally interpreted as $[0,\alpha]$ filter that zeroes all the signals that do not lie in $[-\alpha,0]\cup[0,\alpha]$. Otherwise, specification should be more verbosely $\alpha=[\alpha_1,\alpha_2]$ with $\alpha_2>\alpha_1>0$ so that the filter zeroes all the signals that do not lie in $[-\alpha_2,-\alpha_1]\cup [\alpha_1,\alpha_2]$.  (A) Filter applied on a signal with spikes or any similar signals with $\tilde{\omega}\approx 0$, producing a mean value $\mu$. (B) Filter applied on a signal where wide spatial extent consists of signals with amplitude above $\alpha$, i.e. $\tilde{\omega}>>0$, but still producing the same $\mu$. \textbf{$\alpha$-clamp filter} is similar, except that signals are clamped to a fixed maximum amplitude rather than zeroed.}
\label{fig:fig_filters}
\end{center}
\vskip -0.2in
\end{figure}

 Two types of filters applied are (1) fraction-pass filter and (2) fraction-clamp filter shown in figure \ref{fig:fig_filters}(A, B) . Assume that $f(x)$ is normalized to $[-1,1]$. Denoting the pass filter as $P_{\alpha}[.]$ where $\alpha>0$ and any function as $f(x)$, we define $P_{\alpha}[f(x)]\equiv f_{\alpha}^{(p)} (x)=f(x)$ if $|f(x)|\le\alpha$ and $f_{\alpha}^{(p)}(x)=0$ otherwise. Denoting the clamp filter as $C_\alpha [.]$, define $C_{\alpha} [f(x)]\equiv f_{\alpha}^{(c)}(x)=f(x)$ if $|f(x)|\le\alpha$ and $f^{(c)}_{\alpha} (x)=\pm\alpha$ if $f(x)>\alpha$ and $f(x)<\alpha$ respectively. However, when specified with $[\alpha_1,\alpha_2]$, in the case of pass filter, $f^{(c)}_{(\alpha_1,\alpha_2)} (x)=f(x)$ if $|f(x)|\in[\alpha_1,\alpha_2]$ and zero otherwise. In another words, $\alpha$ is used as an abbreviation to denote more precisely the range $[0,\alpha]$. If $f(x)$ is not normalized, let us denote the normalized function as $f_1 (x)$ where $f(x)=N\times  f_1 (x)$. In this paper, for each $x$, $N=\displaystyle\max_{x} |f(x)|$. Thus, the filter is applied in the following manner: $N\times P_\alpha [f_1 (x)]$.

\par Training is conducted from scratch for different data scale, i.e. each 6-modality image is converted to the following sizes and trained on many different 3D U-Nets. (1) 4x U-Nets in series A are trained on data resized to $(6,19,48,48)$, which we label A1, A2, A3, A4. We will refer to them as series A. (2) Similarly, we train 4x U-Nets in series B with $(6,19,96,96)$. (3) 4x in series C $(6,19,144,144)$ and (4) 4x in series X $(6,19,192,192)$. The baseline for our internal standard is not Dice coefficient performance. Instead, it is less biased to consider both Dice and the fixed number of epochs, $n_{epoch}=80$ , since Dice coefficient can be easily overfit with the small data and large number of epochs. Since this is the first analysis on specific spiking noise behavior, we conduct the most basic training procedure without procedural modification such as data augmentation etc. There is no guarantee that procedural modification might improve the result and thus we leave it for further studies, expecting particular behaviors to each modification (rather than just incremental performance improvement).

\par With batch size=1, X series U-Nets took around 5 hours (other series are significantly faster) and the average Dice score in Pytorch training mode can achieve a competitive value of 0.554 (see annex table 1). Note that recent attempts have shown many overfitting results in the official competition website. Compare the results with \cite{10.1007_978-3-319-55524-9_22}, the top entry of ISLES2016 challenge with Dice coefficients around 0.31 achieved during the competition period. The group also tops the ISLES2017 whose dataset is an improved version of ISLES2016. In our case, while it is possible to achieve Dice score up to 0.77 with further training, we refrain from using these networks that have even greater risk of overfitting (besides longer training time). We trained all the networks on ASPIRE1 provided by NSCC, Singapore, whose specification can be found in \cite{nscc_help}.

\section{Results and Discussions}

\par The idea of amplitude filtering is mainly motivated by raw LRP output exemplified by figure \ref{fig:meim_res3}(A). Since convolutional filters are spatially sensitive, relevance propagation, which is a form of function inverse, will also be spatially sensitive. In fact, the algorithm explicitly uses spatial information of the input that has been forward propagated. We then suppose that undesirable high localized intensity like figure \ref{fig:meim_res3}(A) originates from reinforcement of high amplitude signals propagated in a way that enhance one another cumulatively. By using filters that de-emphasize the importance of high amplitude signals, useful pattern contributed by lower amplitude signals can be uncovered. Without knowing how much signal to filter away exactly, the filtered output still suffers from noises as shown in figure \ref{fig:meim_res3}(B).

\par \textbf{Quantifying the improvement in interpretability.} Since there has not been a standardized measure of AI interpretability for saliency maps, in this paper we define coefficients to measure inclusivity. In particular, we use \textit{input-inclusivity} $D(LRP_\alpha,x)$ and \textit{ground-truth inclusivity} $D(LRP_\alpha ,y_{OT})$ where $LRP_\alpha$ is the output from filtered relevance propagation $LRP_\alpha = LRP_\alpha(y),y=f_{3D U-Net}(x) $ and $y_{OT}$ the ground-truth segmentation. Both $y, y_{OT}$ are binarized.  The inclusivity coefficient of $b$ in $a$ is defined as
\begin{equation}
D(a,b)=|a\cap b|/|b|
\end{equation}
which gives a high value of $1.0$ when $b\subseteq a$ and lower value when their region of intersection is small. The function should also be smoothened by adding a constant to denominator where appropriate. We suggest the following interpretation of LRP output. LRP output with higher \textit{ground-truth inclusivity} (nearer to 1.0) provides more interpretable information, and this happens when LRP output highlight the region with lesion and some other region beyond the lesion that matters to the prediction of lesion. As for \textit{input-inclusivity}, we check that LRP is properly taking into account at least input region with high intensity; if it does, it will score nearer to $1.0$. In essence, we cannot rule out the possibility that regions outside the lesion might actually contain useful information regarding the position and size of a lesion from medical point of view. Hence, the value should be taken as only a first indicator and analyzed along with the inspection of the actual output.

\par
\begin{figure*}[ht]
\vskip 0.2in
\begin{center}
\centerline{\includegraphics[width=\textwidth, trim = {0 0cm 0 0cm}]{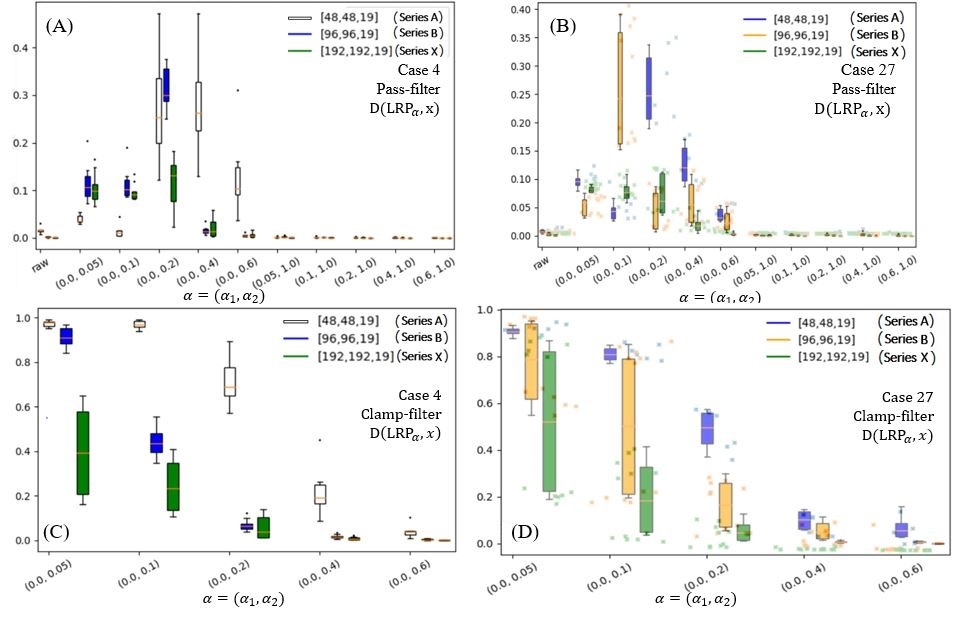}}
\centerline{\includegraphics[width=\textwidth, trim = {0 0cm 0 0.5cm}]{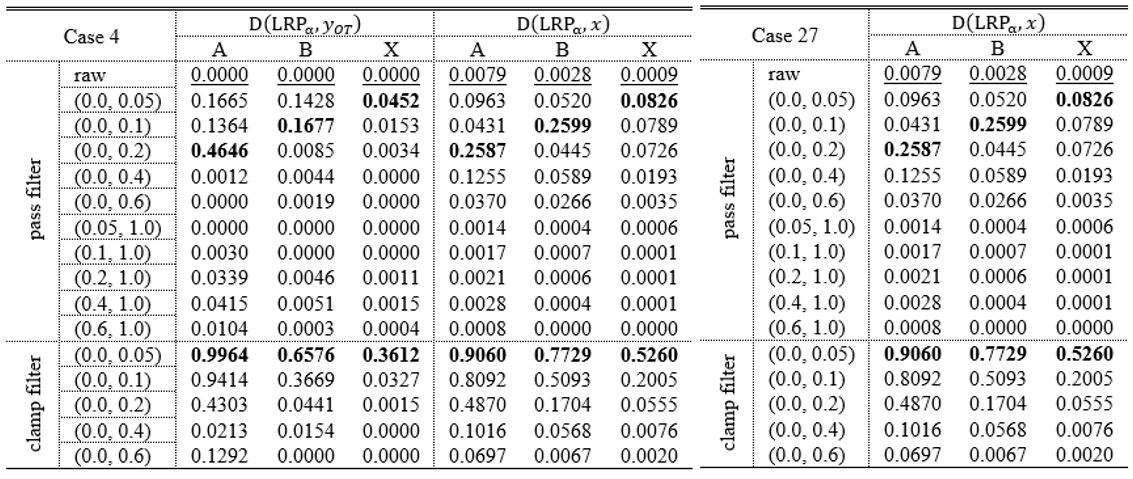}}
\end{center}
\caption{\textbf{Quantification of LRP output interpretability via inclusivity coefficients.} (A, B) $D(LRP_\alpha ,x)$ values for case 4 and 27 respectively, plotted against different values of $\alpha=(\alpha_1,\alpha_2)$. (A) In case 4, for example, $\alpha=(0,0.2)$ for series A is shown in white bar. The value is an improvement from the raw LRP coefficient. The mean value is bolded 0.4646 in the table. (B) In case 27, for series B with $\alpha=(0,0.2)$ blue bar shows improvement from raw LRP coefficient also. The mean value is bolded 0.2599. (C, D) Similar to (A) and (B), but the coefficients are shown for clamp filters. Bolded values show the highest mean value across different filters in the series. In figure (D), $x$ points represent the data points used to plot the box plots in the corresponding colors. Each point is a value of inclusivity coefficient per channel.}
\label{fig:inclusivity_plot}
\vskip -0.2in
\end{figure*}

\begin{figure*}[hp!]
\vskip 0.2in
\begin{center}
\centerline{\includegraphics[width=\textwidth, trim = {0 0.5cm 0 0.5cm}]{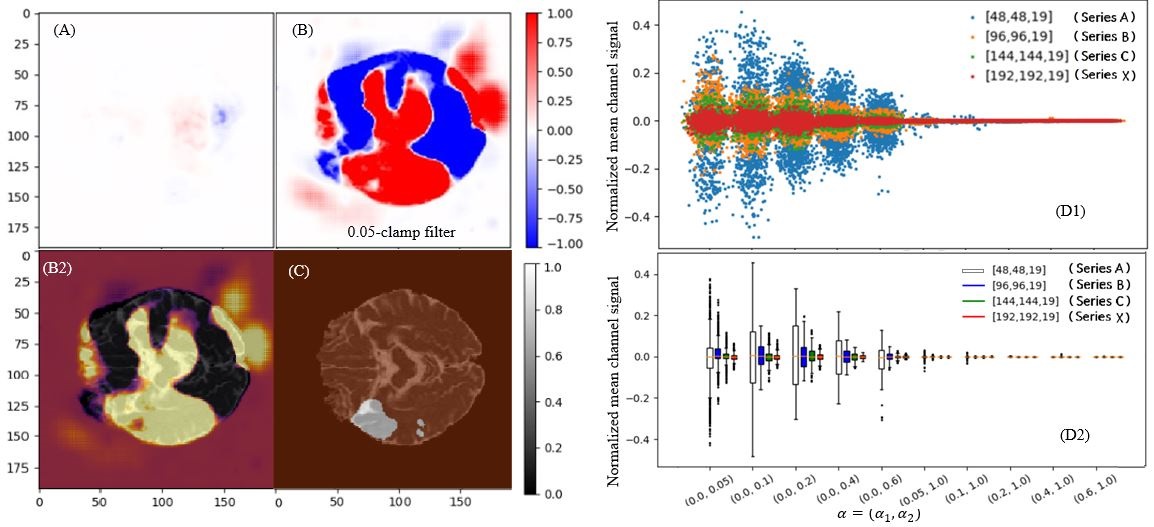}}
\end{center}
\caption{\textbf{$\alpha$-dependency on input scale.} (A-C) LRP output from 3D U-Net X5, showing a slice in the ADC modality of the image of patient with case number 27; see annex table 1. Dice coefficient is $0.755$.  (A) raw LRP output produces highly localized signal. (B, B1) $0.05$-clamp filter applied on LRP, producing signals with some visible artifacts; $max|R_c|=0.00014$. Also see annex figure 4 for $0.05$-pass filter showing unsuccessful extraction. (C)  ground-truth segmentation overlaid on the ADC slice. (D) \textbf{Signal distribution for various $\alpha$-pass filters} showing possible signal distribution trend and a sweet-spot for signal extraction. (D1) Scatter plot: vertical axis shows mean normalized LRP output signals, averaged over the spatial extent of each channel; horizontal axis shows the filter. For a training data point, all 6 channels are plotted in the same figure, also randomly horizontally scattered. Invalid values are not included. (D2) Box plots of D1.}
\label{fig:meim_res3}
\vskip -0.2in
\end{figure*}

 The coefficients are now used to quantify the visual comparison of figure \ref{fig:fig_meim3_main}(A, A2) with figure \ref{fig:fig_meim3_main}(B, B2, C, C2) suggested in the introduction section; for that particular patient whose case number is 4, observe in particular figure \ref{fig:inclusivity_plot}(A) with $\alpha=(0,0.4)$ white bar and figure \ref{fig:inclusivity_plot}(C) with $\alpha=(0,0.2)$ white bar. Now we observe the general trend. From figure \ref{fig:inclusivity_plot}(A, B), raw LRP output are generally scoring low in $D(LRP_\alpha,x)$ inclusivity coefficients. The boxplots show improvements in the distribution of coefficient values when pass filters $\alpha=(0,\alpha_2)$ are applied, i.e. when higher amplitude signals are removed. The table also shows the mean values of the coefficients improved when filters $\alpha=(0,\alpha_2)$ are applied. For example, for series A, for $\alpha=(0.0,0.2)$ the mean value of $D(LRP_\alpha,x)$ improved from 0.0079 (underlined in the table) to 0.2587 (bolded in the table). On the other hand, when low amplitude signals are removed, i.e. when $(\alpha_1,\alpha_2)$ pass filters are used with $0<\alpha_1<\alpha_2$, inclusivity coefficients are nearly uniformly zeros, even with $\alpha_1$ as low as 0.05. This suggests that a large amount of information is stored in low intensity signals.

\par As for clamp filters, as shown in figure \ref{fig:inclusivity_plot}(C,D), improvements are a lot more obvious for all filters tested. In the same vein, the coefficients can be used to quantify the visual comparison for the improvements of interpretability from figure \ref{fig:meim_res3}(A) to figure \ref{fig:meim_res3}(B). This is shown in figure \ref{fig:inclusivity_plot}(B, D) for case 27.

\par \textbf{Finding relevant signals in lower amplitude regime.} From figure \ref{fig:fig_meim3_main}, zeroing the signals whose amplitude range lies in the top $60\%$ amplitude $(\alpha=0.4)$ or clamping the signals in the top $80\%$ $(\alpha=0.2)$ reveals a more sensible LRP output. The heatmap indicates that the region corresponding to the brain slice contributes to the decision of lesion segmentation that closely matches the ground -truth. The figure uses an example from U-Net A1 which involves major resizing to $(19,48,48)$. By contrast, for X2 U-Net (figure \ref{fig:inclusivity_plot}(A-C)), clamping the top $95\%$ of the signals $(\alpha=0.5)$ will yield a heatmap that covers the entire brain slice, but with unwanted artifacts outside the brain slice. When clamping $90\%$ instead $(\alpha=0.1)$, the heatmap stays within the brain slice, but utilizes relatively less information from some other part of the brain (not shown). The output of $\alpha=0.1$ does not necessarily reflect poor performance, since region further away from the lesion may affect the lesion less. Unfortunately, we need more concrete medical knowledge to verify this and to constraint the output to region more consistent with medical science. 

\par A single point in figure \ref{fig:meim_res3}(D1) corresponds to one of the 6 channels from the LRP output. Note that each such channel in LRP output provides the “explanation” on how much regions in the 3D spatial domain contributes to the lesion prediction. This single channel output signal (still in the shape D, H, W) is normalized against the maximum absolute value and then averaged across the spatial domain, giving a point in the figure. We note that the output in different channels typically are close to one another within the same input data. All 6 channels are plotted in the same figure.

\par The result above suggests a sweet-spot $\alpha=(0,\alpha_0)$ that minimizes the appearance of artifact but maximizes the utilization of relevant spatial information. For $\alpha$-pass filter, $\alpha=(0,\alpha_2)$ with $\alpha_2\in [0.1,0.6]$ shows a trend revealing the variability of information stored in the normalized LRP final output; see figure \ref{fig:meim_res3}(D1, D2). We treat this variability as information content and loosely relate it with $\mu$ shown in figure \ref{fig:fig_filters}, and naturally with area under the curve as well. Informally, then, we hypothesize that greater range and variability of $\mu$ (and thus of area under the curve) reflects useful information content

\par From series A, or blue points of figure \ref{fig:meim_res3}(D1), white bars of figure \ref{fig:meim_res3}(D2), when $\alpha_1=0$, we see that lower $\alpha_2$ yields greater mean signal variability. From figure 3, this reflects greater range of $\mu$ that the data points can take. With higher $\alpha_2$ and $\alpha_1=0$, on the other hand, $\mu$ is more constrained. However, when $\alpha=[0,0.05]$, it appears that series A suffers from a loss of variability of $\mu$. Following the previous hypothesis, we might have filtered away information beyond the sweet-spot $\alpha_0$ and lose all the useful information. Annex figure 2 shows similar trend as figure \ref{fig:meim_res3}(D1) for clamp filter, except it is better-behaved. The clamped amplitude still can contribute to variability. Although the mechanism is also unknown, we suggest that, for clamp filters, very small value $\alpha$ might be necessary to attain the sweet-spot.

\par \textbf{Propagation of only-high-amplitude signals.} As mentioned briefly in the introduction, correct prediction does not imply correct use of input information. Figure \ref{fig:fig_meim3_main}(A) shows local high intensity (spikes) nearer the top-left of the figure, despite good Dice score. Figure \ref{fig:meim_res3}(A) shows even more extreme signals accumulation. This suggests the optimization has reached the stage of false minima. We either need further training using the same data (risking overfitting), apply image transformations or provide larger dataset with greater varieties in the samples.

\par Figure \ref{fig:meim_res3}(D1, D2) shows information variability $\mu$ quickly decays when $\alpha=[\alpha_1,\alpha_2]$ with $\alpha_1>0$. Even with $\alpha_1 = 0.05$, i.e. by ignoring only the lowest $5\%$ signal, we lose nearly all range of $\mu$. If the signals are always zero, then this is expected. However, we do have non-zero signals (annex figure 3), implying that high-amplitude signals are transmitted in spikes. In general, this leads to observation 3 in the introduction that relatively low magnitude signals propagated in a NN stores significant amount of information.

\par \textbf{Presenting interpretability improvement selectively}. As a reminder, training is done in 80 epochs and not aimed at maximizing Dice score (avoid overfitting). Thus, interpretability improvements are shown only for cases whose predicted lesions gave good Dice score, such as case 4 and 27. For poor Dice score, i.e. bad predicted output, high interpretability value will be meaningless. Interestingly but perhaps not surprisingly, a patient case tends to score similarly across different U-Net trained. Case 4 and 27 score well. Case 2 and 45 tend to score poorly (small, scattered lesion, not shown).

\section{Where to Go from Here}
\par \textbf{Studying sub-optimality of deep neural network (DNN) to understand its inner mechanism.} Sometimes, it is not clear when to stop the training. When high training or validation accuracies are attained, sometimes a longer training time does improve the accuracy incrementally, but we cannot be sure of the usefulness of such apparent improvement. Neither can we be sure if a large amount of weights actually changed during this process (in case of training with stochastic element). This opens up a possibility of research where we monitor the evolution of interpretability signal as a DNN is being trained and provide an evaluation of the DNN different from the usual metric such as accuracy. The following benefits can be envisioned:
\begin{itemize}
\item A procedural modification to the training process can thus be justified according to the result from interpretability study.
\item New independent performance metrics can be designed to evaluate a DNN. This independence may be achieved because interpretability method such as LRP does not interfere with training procedure.
\end{itemize}

\par \textbf{Investigating normalized signals.} We have compared normalized signal values so far, where the signals are normalized by absolute maximum value per channel of a single 6-modality image, discarding the amplitude information. Now we justify this choice. For spatially dependent neural network such as convolutional networks and U-Net, amplitude of the feature maps may vary with the slightest variation of the network. For example, a larger kernel size of convolutional filter will already sum up greater number of values from the previous layer per iteration. Averaging can certainly be done but it is neither clear how loss optimization affects this nor how layers such as batch normalization regulates forward, backward and relevance propagations. Unfortunately, this is not trivial. As an example, Kaiming initialization \cite{He_2015_ICCV} does affect the performance of a network. From relevance propagation formula in equation (\ref{eqn:lrpformula}), we do see how signals are summed without magnitude normalization factors. We distinguish magnitude normalization factors from content-specific weights $w_{ij}$ values in equation (\ref{eqn:lrpformula}). An example of magnitude normalization factors is the value $1/9$ for convolutional kernel $\frac{1}{9} K$ with 3x3 kernel, where $K$ can be 3x3 matrix with all entries being 1.

\par \textbf{Comparing the same layer in different U-Net.} Figure \ref{fig:fig_filters} can be analyzed by fixing $\mu$. Comparing figure \ref{fig:fig_filters}(A) and \ref{fig:fig_filters}(B), a variation of signals behavior can produce the same $\mu$. When $\tilde{\omega}\approx 0$, spike-like signals are more likely to occur. On the other hand, when $\tilde{\omega}>>0$ the intermediate 3D feature maps produced via relevance propagations contain larger region with high signal amplitude. To compare the performance of 2 different networks, we might thus observe the amplitude of the same layers and possibly create correlation between signal amplitudes (which we have not utilized) and network’s performance. 

\par \textbf{Filter Calculus.} Let us introduce a semi-abstract LRP filter calculus to formalize some of the observations. Let $f(x)$ be the lesion segmentation output, where $f(.)$ is the 3D U-Net and $x$ a multi-modal image. Let $R_L=y$ and $R_n=relprop(R_{n+1})$ denotes relevance propagation, corresponding to the appropriate variant of equation (\ref{eqn:lrpformula}), and $R=R_0=f_{lrp}^\alpha (R_L)$ be the LRP output under a filter with $0<\alpha_0<\alpha\le 1$. Let $R^c$ be channel $c$ of $R$. Figure \ref{fig:meim_res3}(D1) suggests the following hypothesis. Given the spatial index $w\in W$ of an input image, there exists $\mu<1$ such that 
\begin{equation}
\langle W_c^\alpha \rangle\equiv \int_{w\in W} w\times N_{abs} \big[f^\alpha_{lrp}[R_L (w)]\big]_c dw<\mu^\alpha
\end{equation}
where $N_{abs} (x)=x/max|x|$.

\par However, considering $(0,0.6)$-pass filter from figure \ref{fig:meim_res3}(D1) and some outlier-looking values, it might be more precise to frame it in the following. There exist $\mu^\alpha <1$ and probability $0<<p_\alpha<1$ such that $P(\langle W_c^\alpha\rangle\le \mu^\alpha)\ge p_\alpha$ . Using area under the curve $A_I$ instead of $\mu$, we can also similarly suggest there exist $A_I^\alpha<1_W, p_\alpha>>0$ so that $P(A[W_c^\alpha ]\le A_I^\alpha )\ge p_\alpha$ where $A[W_c^\alpha ]\equiv\int_{w\in W} dw (f_{lrp}^\alpha [R_L (w)])_c$ with $1_W$ is some maximum value in the language of the system. This formalizes the $\alpha$-dependent variability we used in the previous section as the quantity $\mu$ or $A_I$.

\par We want to abstract the above hypothesis. Denote the set of filters $\bar{F}$ such that $P_\alpha [.],C_\alpha [.]\in\bar{F}$ and any $F_\alpha\in\bar{F}$ has the abstract properties that generalize $P_\alpha [.],C_\alpha [.]$ and $0<\alpha_0\le\alpha\le 1$. Given a sequence of functions $f_i$ and filters $F_{\alpha,i}\in F$ so that $f_\alpha (x)=(\Pi_i F_{\alpha,i} f_i )(x)$ and a set $W\in\bar{W}$ where $\Pi$ is used to denote repeated application of function composition. We leave the specification of abstract properties for further work, when more analysis and survey may include a greater range of possible functions. Treating the probabilistic $\alpha$-dependent variability as the fundamental structure (and arbitrarily uses the area under the curve representation rather than mean value representation), we write for all $W\in\bar{W}$ the following. Given $A_\alpha\equiv\int_W f_\alpha (w)dw\le 1_W$, there exists $A_I^\alpha$ and $p_\alpha\in (0,1]$ such that $P(A_\alpha \le A_I^\alpha )\ge p_\alpha$.

\par Now we show the approximation steps. Given $y=G(x)=(\Pi G_i )(x)$; this function can be for example a neural network. For each $G_i$, define relevance-propagation by $g_i$ , although generally it is not necessary to have one-to-one correspondence between $g_i$ and $G_i$. Denote $x_p=g(y)=(\Pi g_i )(y)$ as the “true relevance” or “true explanation”. If we have only a noisy version of $G$ denoted $y_1=G_1 (x)$, then $g(y_1)$ gives an inaccurate explanation $x_{p,1}$. The function $G_1$ can be, for example, a neural network not yet well-trained. To recover the true explanation $x_p$, we determine for each explanation a filter, i.e. we define the ordered-pair $(F_{\alpha,i},g_i)$ with $x_{p,\alpha}\approx g_\alpha (y)=(\Pi F_{\alpha,i} g_i )(x)$ that fulfills the abstract properties not yet fully specified. Finally, find the optimal $\alpha=\alpha_0$ so that $x_{p,\alpha}\rightarrow x_p$.

\par We give a trivial example using 2 pass-filters and 2 functions as the following. Here, we use integration as in the usual real number integration. Given unspecified $y=G(x)$ whose explanation is given by $g(x)$. In the main body of this paper,$G(x)$ represents 3D U-Net and $g(x)$ will correspond to LRP process. Then, for a noisy version of $G(x)$, explanation $g(x)$ is to be approximated by $(F_{\alpha,i},g_i )$ sequence.  Let $\bar{W} \equiv{[0,1]}$, $F_(\alpha,1)=F_(\alpha,2)=P_\alpha$, $g_1 (x)=g_2 (x)=x$ for $x\in [0,1]$ and $1_W=1$. We show that this system fulfills the structure above. First, we see that $g_\alpha (x)=F_(\alpha,2) g_2 F_(\alpha,1) g_1 (x)=P_\alpha (x)$.  Then $A_\alpha=\int_0^1 [P_\alpha (x)dx=\int_0^\alpha x dx=\alpha^2 /2$. Indeed, with the choice $A_I^\alpha=\alpha^2,p_\alpha=1$, we see that $P(A_\alpha=\alpha^2 /2\ge A_I=\alpha^2 )=1\ge p_\alpha=1$.

\par Recall that we have $\alpha_0$ sweet-spot value to delineate the extent of information we can filter away to recover useful information. The task of studying this system could be for example finding $\alpha_0$ to recover a function whose inverse is approximated by $g_\alpha$ best or even design the appropriate $(F_{\alpha,i},g_i)$ sequence. We extend the trivial example from above. Suppose $f_\alpha$ has been deployed to approximate a true explanation $g(x)= P_{[0.1,0.5]}  (x)$, which is an $\alpha$-pass filter with $\alpha=[0.1,0.5]$. If $G(x)$ is not corrupted (for neural network, it means it is perfectly trained if such NN exists), we can easily get $\alpha_0$. Up to $\alpha_0=0.1$, the function will still be correctly recovered. Below it, we will get errors.

\section{Conclusion}
\par We have provided LRP implementation on a medical segmentation problem, the first we are aware of. The output is often suboptimal, with LRP showing heatmap highlighting obscure, localized areas or negative contribution to otherwise accurate segmentation. Filters are applied to “extract” interpretable information from high amplitude signals we suspect have masked away useful signals as they reinforce themselves through the layers. A mathematical abstraction in the form of filter calculus is also introduced as a possible method to perform rectification to noise corrupted functions and function approximation. The LRP output is not optimal, partially due to the small relevant dataset used, even though Dice coefficient indicates competitive performance. Clearly, good accuracy alone does not guarantee that a neural network is “thinking” in the right way. Furthermore, more robust medical knowledge is certainly required to constraint LRP output per-se. 

\clearpage
\section*{Acknowledgment}
This research was supported by Alibaba Group Holding Limited, DAMO Academy, Health-AI division under Alibaba-NTU Talent Program. The program is the collaboration between Alibaba and Nanyang Technological university, Singapore. We also thank NSCC for providing high performance computational resources. We also thank Mane Ravikiran Tanaji for the help in setting up the NSCC system for our access.

\bibliographystyle{unsrtnat}
\bibliography{meimLRP}

\end{document}


\begin{figure}[ht]
\vskip 0.2in
\begin{center}
\centerline{\includegraphics[width=\columnwidth, trim = {0 0.5cm 0 0.5cm}]{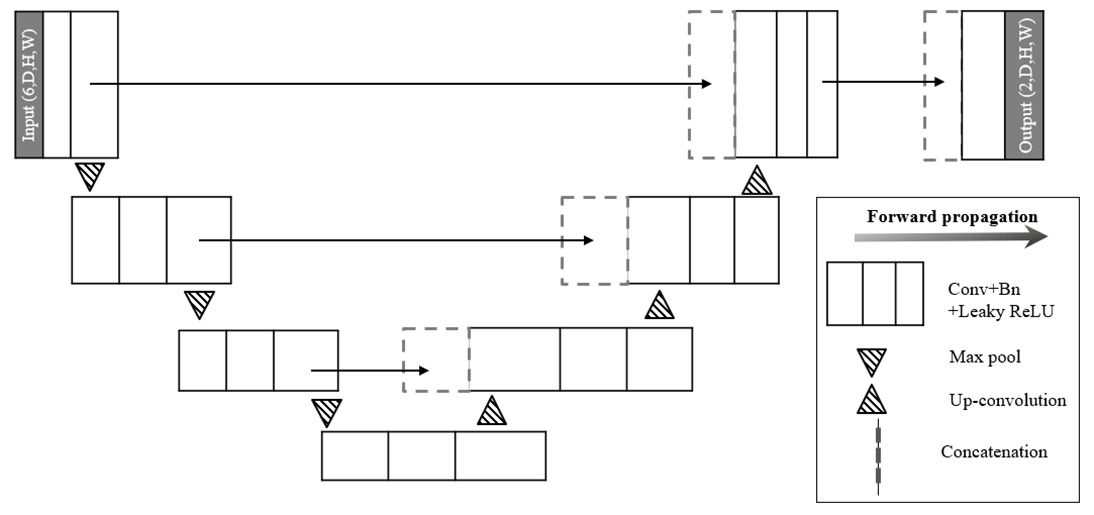}}
\end{center}
\caption{Standard U-Net with slight modification.}
\label{fig:fig_meim3_unet}
\vskip -0.2in
\end{figure}

\begin{figure*}[ht]
\vskip 0.2in
\begin{center}
\centerline{\includegraphics[width=\textwidth, trim = {0 0.5cm 0 0.5cm}]{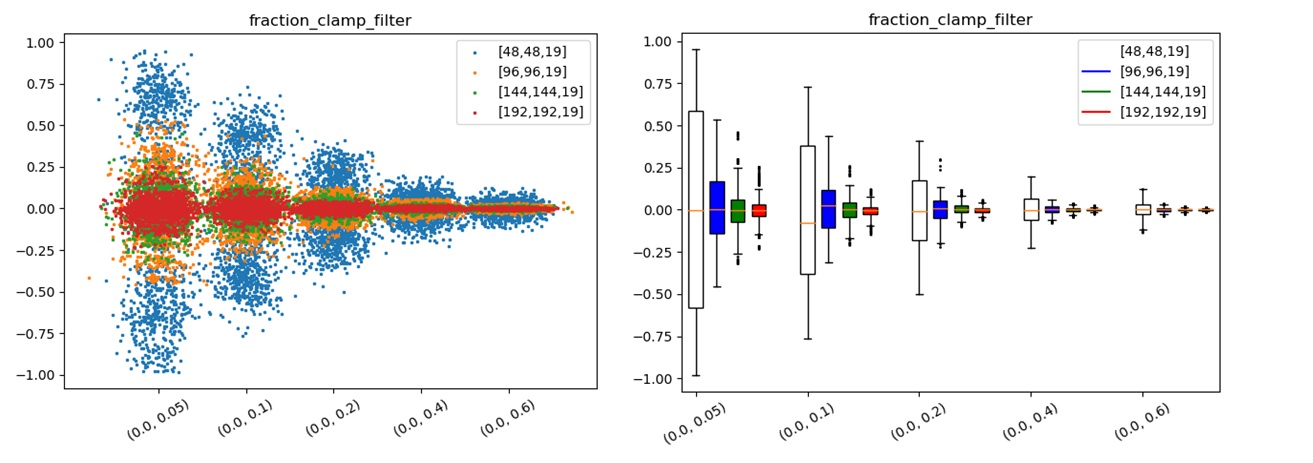}}
\end{center}
\caption{Similar to figure 5(D1, D2), but for $\alpha$-clamp filters. For $(\alpha_1,\alpha_2 )$, we note that with even small $\alpha_1>0$, the result becomes highly unstable. The magnitude appears to accumulate very quickly.}
\label{fig:annex_fig2_Calphatrend}
\vskip -0.2in
\end{figure*}

\begin{figure}[ht]
\vskip 0.2in
\begin{center}
\centerline{\includegraphics[width=0.6\columnwidth, trim = {0 0.5cm 0 0.5cm}]{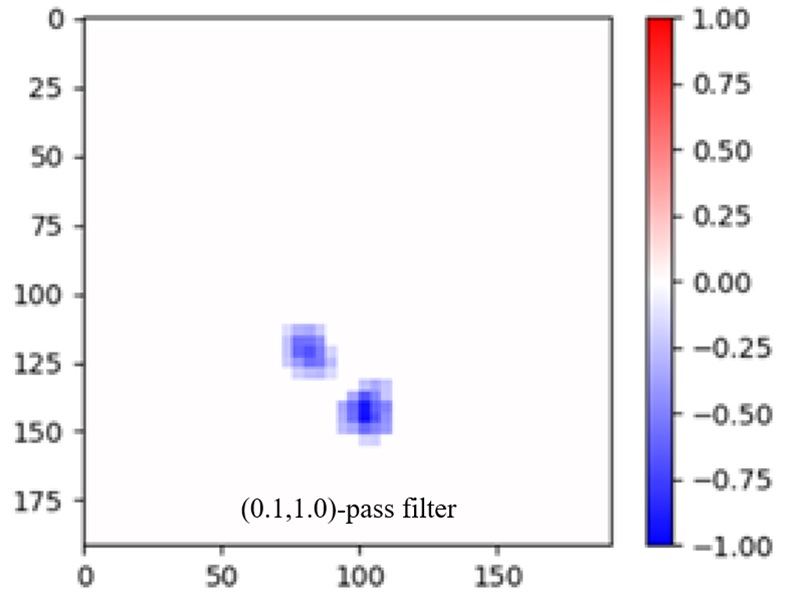}}
\end{center}
\caption{$(0.1,1.0)$-pass filter applied on LRP, producing spikes. Patient case number 4, 3D U-Net A, a slice from rCBF modality, indicating $\tilde{\omega}\approx 0$ propagation of signals.}
\label{fig:annex_fig3_spikes}
\vskip -0.2in
\end{figure}

\begin{figure}[ht]
\vskip 0.2in
\begin{center}
\centerline{\includegraphics[width=0.6\columnwidth, trim = {0 0.5cm 0 0.5cm}]{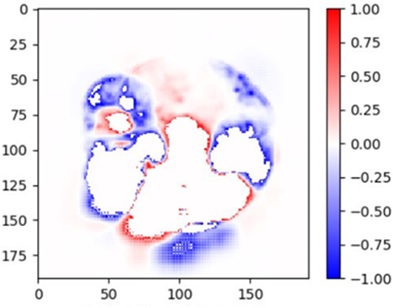}}
\end{center}
\caption{$0.05$-pass filter applied to LRP on 3D U-Net X5 for case number 27 ADC modality.}
\label{fig:annex_fig4_artifacts}
\vskip -0.2in
\end{figure}

\clearpage

\begin{algorithm}[tb]
\caption{Relevance propagation for each applicable layer. Forward propagation is denoted forward(). Gradient propagation is denoted gradprop().  Buffer $\epsilon$ prevents division by 0. self.X is the input to the layer stored as forward propagation is performed before Relprop(). $\epsilon=10^{-9}$. Multiplication and division are element-wise. }
   \label{alg:lrp}
\begin{algorithmic}
\STATE \textbf{function} Relprop(R):
\STATE \hskip1.0em $Z = forward(self.X)$
\STATE \hskip1.0em $S = R/(Z + \epsilon)$
\STATE \hskip1.0em $C = gradprop(S)$
\STATE \hskip1.0em $R = X*C$
\STATE \hskip1.0em return R
\end{algorithmic}
\end{algorithm}

\begin{table}[t]
\vskip 0.15in
\begin{center}
\begin{small}
\begin{sc}
\begin{tabular}{lcccr}
\toprule
 & Time [min] & Time [hour] & $<Dice>$ \\
\midrule
X1	& 296.01	& 4.93	& 0.443039394 \\
X2	& 294.86	& 4.91	& 0.305574958 \\
X3	& 301.76	& 5.03	& 0.00311809* \\
X4	& 296.50	& 4.94	& 0.302716623 \\
X5	& 294.40	& 4.91	& 0.449010816 \\
X6	& 296.79	& 4.95	& 0.213563865 \\
C1	& 151.03	& 2.52	& 0.388121947 \\
C2	& 154.66	& 2.58	& 0.225427605 \\
C3	& 154.24	& 2.57	& 0.254080092 \\
C4	& 151.56	& 2.53	& 0.347355277 \\
B1	& 72.26	& 1.20	& 0.553999599 \\
B2	& 72.32	& 1.21	& 0.391368908 \\
B3	& 71.06	& 1.18	& 0.368551397 \\
B4	& 71.01	& 1.18	& 0.212295298 \\
A1	& 23.29	& 0.39	& 0.324461827 \\
A2	& 21.28	& 0.35	& 0.273856772 \\
A3	& 21.23	& 0.35	& 0.500503654 \\
A4	& 21.26	& 0.35	& 0.326139371 \\
\bottomrule
\end{tabular}
\end{sc}
\end{small}
\end{center}

\caption{Different 3D U-Nets trained from scratch. The code A, B, C and X correspond to the sizes to which the multi-modal images are converted before training. * indicates mode collapsed network.}
\label{table:annex_unet}
\vskip -0.1in
\end{table}